# Structural, morphological, and magnetic characterizations of $(Fe_{0.25}Mn_{0.75})_2O_3$ nanocrystals: a comprehensive stoichiometric determination

John C. Mantilla[a,b*], Luiz C. C. M. Nagamine[a*], Daniel R. Cornejo[a], Renato Cohen[a], Wesley de Oliveira[c], Paulo E. N. Souza[b], Sebastião W. da Silva[b], Fermin F. H. Aragón[b], Pedro L. Gastelois[d]; Paulo C. Morais[b,e], José A. H. Coaquira[b]

[a]Universidade de São Paulo, Instituto de Física, São Paulo SP 05508-090, Brazil
[b]Universidade de Brasília, Instituto de Física, Núcleo de Física Aplicada, Brasília DF 70910-900, Brazil
[c]Universidade Federal de Mato Grosso do Sul, Instituto de Física, Campo Grande MS 79070-900, Brazil
[d]Centro de Desenvolvimento da Tecnologia Nuclear – CDTN, Av. Antônio Carlos, 6627, Pampulha, Belo Horizonte MG 31270-90, Brazil
[e]Universidade Católica de Brasília, Programa de Pós-Graduação em Ciências Genômicas e Biotecnologia, Brasília DF 70790-160, Brazil

**ABSTRACT**

This study focuses on the structure and magnetism of a nanostructured compound, initially labeled as $FeMnO_3$, prepared by the sol-gel method. Through Mössbauer spectroscopy analysis and Rietveld refinement of X-ray diffraction data, the composition of the compound was well determined: a majority bixbyite phase (86 mol%, 94 wt%) with $(Fe_{0.25}Mn_{0.75})_2O_3$ stoichiometry and average crystallite size of ~48 nm, plus a minority hematite phase (14 mol%, i.e., 6 wt%) with an average crystallite size of ~8 nm.
The Raman spectrum exhibits characteristic vibrational bands at 659 $cm^{-1}$, 519 $cm^{-1}$ and 416 $cm^{-1}$, confirming the majority phase, with no features associated with the minority phase. X-ray photoelectron spectroscopy analysis confirmed the presence of oxygen vacancy onto the $(Fe_{0.25}Mn_{0.75})_2O_3$ particle surface, with varying oxidation states ($Fe^{3+}$, $Fe^{2+}$, $Mn^{3+}$, and $Mn^{4+}$). X-band magnetic resonance data revealed a strong and broad resonance line in the whole temperature range (4.3 K ≤ T ≤ 300 K), dominated by the majority phase, with *g*-value decreasing monotonically from (2.93 ± 0.01) at 50 K down to (2.18 ± 0.01) at 300 K. The temperature dependence of both resonance field and resonance linewidth shows a remarkable change in the range of 40-50 K, herein credited to surface spin-glass behavior. The model picture used to explain the MR data in the lower temperature range (below about 50 K) assumes $(Fe_{0.25}Mn_{0.75})_2O_3$ nanoparticles with a core-shell structure. Results indicate that below about 50 K the shell's spin system reveals a paramagnetic to spin-glass-like transition upon cooling, with a critical temperature estimated at (43 ± 1) K. In the higher temperature range (above ~50 K), the superparamagnetic minority phase contributes remarkably to the temperature dependence of the resonance linewidth. Zero-field-cooled and field-cooled data show strong irreversibility and a peak in the ZFC curve at ~33 K, attributed to a paramagnetic-ferrimagnetic transition of the majority phase.

\* Corresponding Author:  mantilla52@gmail.com; nagamine@if.usp.br

# 1. INTRODUCTION

Nanotechnology has engendered a paradigm shift across various domains of science and technology, empowering manipulation and governance of matter at the nanoscale. The extraordinary and distinct properties of nanomaterials have ushered in novel prospects for applications in medicine, electronics, energy, and a plethora of other fields. Moreover, nanotechnology has yielded substantial progress in the fabrication of more efficient and miniaturized devices, fostering a world that is increasingly interconnected and technologically sophisticated. Given its capacity to tackle intricate challenges and enhance the quality of life, nanotechnology continues to stand as a captivating and auspicious arena for research and development endeavors.

Nanostructured magnetic semiconductors, including mixed metal oxides, hold particular interest for technological applications due to the versatile nature of their physical and chemical properties, derived from both stoichiometry and particle size. As widely acknowledged, these properties often differ from those observed in the same materials but with micrometric and larger particle sizes. Among this extensive array of nanomaterials, manganites, which are perovskite oxides with the general formula $R_{1-x}M_xMnO_3$, where R is normally a trivalent rare earth ion and M is usually a transition metal or a divalent alkaline ion, are widely recognized for exhibiting intriguing properties. Materials from the $R_{1-x}M_xMnO_3$ family, especially when the particle size is in the nanometer range, can exhibit remarkable densities of ionic and electronic defects, making them important candidates for the development of solid-state fuel cells, chemical gas sensors, magnetic refrigeration, and eventually as memory storage devices [1-3]. Just as in all perovskites, the composition and structural parameters strongly influence the physico-chemical properties of manganites [4]. On the other hand, it is well established that the fundamental state of a given manganite can be modified by variations in basic thermodynamic variables such as pressure and strain, as well as by the application of electric and magnetic fields [5-8]. The basic building block of manganites with a perovskite structure is the $MnO_6$ octahedron. According to Travis *et al.* [9], a compound from the manganite family will exhibit a stable perovskite structure when the Goldschmidt tolerance factor *t* is within the range of $0.89 < t < 1.02$. Considering a particular magnetite with composition $AMnO_3$ (where A represents a rare earth ion or a transition metal ion), the Goldschmidt tolerance factor is defined as follows:

$$t = \frac{1}{\sqrt{2}} \frac{r_A + r_O}{(r_{Mn} + r_O)}$$

being $r_A$ ($r_{Mn}$) the ionic radius of the $A^+$ ($Mn^+$) cation and $r_O$ the ionic radius of the $O^-$ anion. Note that $t = 1$ corresponds to the ideal perovskite structure [9]. The perovskite lattice undergoes structural deformation primarily due to two types of distortions: one results from the tilting of the $MnO_6$ octahedron, whereas the other arises from the asymmetry in the six Mn-O bond lengths surrounding the Mn atom within the $MnO_6$ octahedron [10] (the Jahn-Teller distortion [11]). Simultaneously, there has been a remarkable surge in interest towards investigating non-perovskite structures of the $ABO_3$ type (where A and B represent cations such as Co, Ni, Fe, Mn, etc.). This heightened interest can be attributed to their multifunctional properties and promising potential for diverse device applications [12,13]. Notably, manganese oxide has garnered extensive attention due to its intriguing magnetic and multiferroic characteristics [14]. Manganese exists in various valence states, thereby endowing the material with a diversity of magnetic properties contingent on its composition. Furthermore, altering the composition results in concurrent changes in structure and valence states. For instance, in the case of MnO, which adopts a halite-type cubic structure, the Mn ion exhibits a valence state of 2+. In contrast, for $Mn_2O_3$, with a bixbyite structure (either cubic or orthorhombic), the valence state is 3+. Similarly, in $Mn_3O_4$, possessing a spinel cubic structure, the Mn ion displays a mixed valence of 2+ and 3+ [15]. Remarkably, $AMnO_3$-type compounds, including the bixbyite compound $FeMnO_3$, have found versatile applications, such as in negative temperature coefficient (NTC) thermistors, oxidation catalysis, and superparamagnetic materials. Cao *et al.* [16] extensively investigated $FeMnO_3$ as an anode material for lithium-ion batteries. This compound has gained prominence in electrochemistry owing to its intriguing redox behavior, superior theoretical capacity for lithium-ion batteries (500–1000 mAh/g), and lower operating potential. Furthermore, $FeMnO_3$ has been effectively employed in microwave devices and catalysts. The bixbyite-based $(Mn^{3+}, Fe^{3+})_2O_3$ compound adopts the $\beta$-$Mn_2O_3$ crystal structure (cubic, *Ia-3*, *a* = 9.41 Å) [17]. It comprises two metal sites situated at the 24d and 8b crystallographic positions, whereas the oxygens occupy the 48e positions. Both cations exhibit octahedral coordination with oxygen, with the 8b site displaying a slight trigonal distortion, whereas the 24d site exhibits a more substantial distortion. Rayaprol *et al.* [18,19] conducted a thorough investigation of the magnetic and magnetocaloric properties of FeMnO$_3$ synthesized through mechano-synthesis. Their findings reveal that FeMnO$_3$ exhibits ferrimagnetic order at room temperature but undergoes a phase transition to the antiferromagnetic state at 36 K. Below approximately 150 K, the magnetization exhibits a pronounced deviation from Curie-Weiss behavior. It would be anticipated that Fe$^{3+}$ and Mn$^{3+}$ ions, octahedrally coordinated to oxygen, would possess high-spin configurations with quenched orbital moments, leading to magnetic moments of 5.9 μ$_B$ (Fe$^{3+}$) and 4.9 μ$_B$ (Mn$^{3+}$), resulting in an expected average of 5.5 μ$_B$ per metal site. Through fitting the magnetization data in the 200–380 K range using the Curie-Weiss law, the authors previously derived an average magnetic moment per site of 4.1 μ$_B$, suggesting the presence of some degree of antiferromagnetic correlation even at higher temperatures [19]. Various preparation techniques have been successfully employed to synthesize nanosized FeMnO$_3$, including physical methods such as high-energy ball milling, as well as chemical techniques like chemical co-precipitation, sol-gel, and combustion [20-21]. The physical and chemical attributes of the materials produced through the sol-gel approach, such as particle size, surface area, and mechanical properties, can exhibit significant variations depending on factors such as working temperature, operating conditions, and chemical precursors utilized. However, it has been documented that the sol-gel method offers a highly reproducible means of fabricating nanomaterials with elevated surface area and improved mechanical properties when compared to alternative synthetic routes [22]. Nonetheless, it is noteworthy that a limitation of the sol-gel method lies in the sensitivity of the precursor's hydrolysis to water addition. Even under vigorous stirring, the rate of hydrolysis is so rapid that particles tend to precipitate immediately upon water introduction into the reaction medium. A reduced rate of hydrolysis, however, can result in particle size reduction and an increase in surface area, aspects of great interest in catalysis [23]. The stability of various compounds formed within the FeMnO$_3$ system is contingent not only on the Fe/Mn/O ratios but also on the preparation method and calcination temperature [24]. Among the identified compounds are mixed oxides (Fe$_{1-x}$Mn$_x$)$_{1-y}$O, which have been utilized in water-splitting reactions [25], rock-salt oxides Fe$_x$Mn$_{1-x}$O [26], defect spinel γ-FeMnO$_3$ analogous to γ-Fe$_2$O$_3$ [27], and manganese-substituted magnetite Mn$_x$Fe$_{3-x}$O$_4$ [28]. In a study conducted by Ponce *et al*. [29], it was revealed that the stability of Mn$^{4+}$ ions play a critical role in determining the catalytic activity of manganites for methane oxidation in the temperature range of 200-800 °C. Generally, high surface area is associated with small particle sizes, leading to a larger surface area exposed to gases for a given mass of nanoparticles [30]. Morphology control stands as a paramount requisite for enhancing catalyst activity. As anticipated, nanosized $FeMnO_3$ exhibits behavior characteristic of a single magnetic domain, and its magnetic properties are significantly influenced by factors such as size, shape, stoichiometry, inversion parameter, crystallinity, and surface termination. These aspects can be effectively manipulated through preparation and post-treatment synthetic methodologies [31].

Considering the $FeMnO_3$ compound as paradigmatic for this family, there are few studies that have explored the physical and morphological characteristics of compounds with excess or deficiency of Fe, with the exception of the first-principles study by Bazhenova and Honkala [32].

In this investigation, we present a comprehensive exploration of the physical properties exhibited by $Fe_{0,5}Mn_{1,5}O_3$ nanoparticles synthesized using the sol-gel method. A meticulous evaluation of the structural, morphological, optical, and magnetic characteristics was undertaken through a combination of cutting-edge techniques, including x-ray diffraction (XRD), x-ray photoelectron spectroscopy (XPS), transmission electron microscopy (TEM), Raman spectroscopy, Mössbauer spectroscopy (MS), magnetometry, and electron magnetic resonance (EMR).

The proficient integration of MS, XPS and XRD techniques, the latter skillfully resolved through Rietveld analysis, enabled a meticulous elucidation of the system's precise stoichiometry. Additionally, this approach facilitated the unequivocal identification of the distinct phases present in the sample, their corresponding fractions, and the mean sizes of their crystallites. The magnetic properties of the compound were exhaustively characterized using the aforementioned techniques, and the results were effectively correlated with the morphology and composition of the studied material.

## 2. EXPERIMENTAL DETAILS

### 2.1 Sample preparation

The $(Fe_{0.25}Mn_{0.75})_2O_3$ nanoparticles (NPs) reported here were synthesized using sol-gel polymerization method, which utilized transition metal nitrates as precursors. The process involved dissolve manganese nitrate tetrahydrate $(Mn(NO_3)_2·4H_2O)$, ferric nitrate nonahydrate $(Fe(NO_3)_3·9H_2O)$ and citric acid $(C_6H_8O_7)$ in a small quantity of double distilled water. Aqueous solutions of manganese (0.4 mol/L) and iron (0.8 mol/L) nitrates were mixed in stoichiometric proportion ($Fe^{3+}$:$Mn^{2+}$::1:1). The mixture was then diluted in 50 mL of ethylene glycol (99% purity) while keeping the 1:1 volume ratio. The homogeneous reaction medium was stirred at 80 °C to ensure uniformity. Once the gel was formed the temperature was raised to 250 °C for self-ignition reaction. The resulting product was subsequently calcined under ambient atmosphere (900 °C, 72 hours) to obtain the nanosized compound.

**2.2 Characterization Details**

Structural analysis of the as-synthesized powder at room temperature were performed using a commercial XRD diffractometer (Rigaku, model D/max) equipped with copper radiation source Cu-$K\alpha$ ($\lambda$=1.5418 Å) and operating in steps ($2\theta$) of 0.05°. Crystal structures were refined using the Rietveld method [33] with GSASII suite program. Representative TEM micrographs of the sample, including energy dispersive x-ray spectroscopy (EDX), were obtained using a commercial microscope (JEOL, model JEM-2100). Raman spectra were recorded using a triple spectrometer (Jobin-Yvon, model T64000) equipped with a 2048×512 pixels nitrogen-cooled CCD (Charge-Coupled Device) camera. Compositional XPS analysis was performed on a Specs surface analysis system equipped with a Phoibos 150 electron analyzer using a monochromatized Al K$\alpha$ radiation (1486.6 eV) at a power of 350 W. Casa XPS software was used to process the recorded data and to estimate the sample surface atomic concentration. C -1s signal (284.6 eV) was used as reference for calibration of the binding energies (BE) of different elements. Electron magnetic resonance (EMR) spectra (from 3.8 to 300 K) were collected using a commercial X-band spectrometer (Bruker, model EMX) equipped with a rectangular cavity. $^{57}$Fe Mössbauer spectra in transmission mode were recorded at 80 K and 300 K using a conventional spectrometer. The sample was mixed with boron nitrate (reaching 0.1 mg $^{57}$Fe per cm$^2$) and homogeneously dispersed within a nylon-based sample holder. The spectrometer was equipped with a 25 mCi $^{57}$Co source, immersed in Rh matrix. The Mössbauer source was coupled to the driver (room temperature) while a sinusoidal velocity driver was used. A cryostat (Oxford Cryosystems) equipped with a temperature controller was able to keep the sample in the desirable temperature. The dc magnetic measurements were performed using a commercial SQUID (MPMS, Quantum

Design) magnetometer varying the temperature in the range of 5 to 300 K while applying magnetic fields up to ± 70 kOe.

## 3. RESULTS AND DISCUSSION

### 3.1 Mössbauer Spectroscopy

The Mössbauer spectra recorded at 295 K and 80 K and their fits, including the subspectra (colored solid lines), are shown in Figs. 1a and 1b, respectively. These fits were performed using two doublets corresponding to 24d and 8b crystallographic positions of the (Fe,Mn)$_2$O$_3$ phase, plus two sextets identified as hematite ($\alpha$-Fe$_2$O$_3$), and amorphous hematite. The obtained hyperfine parameters are collected in Table 1.

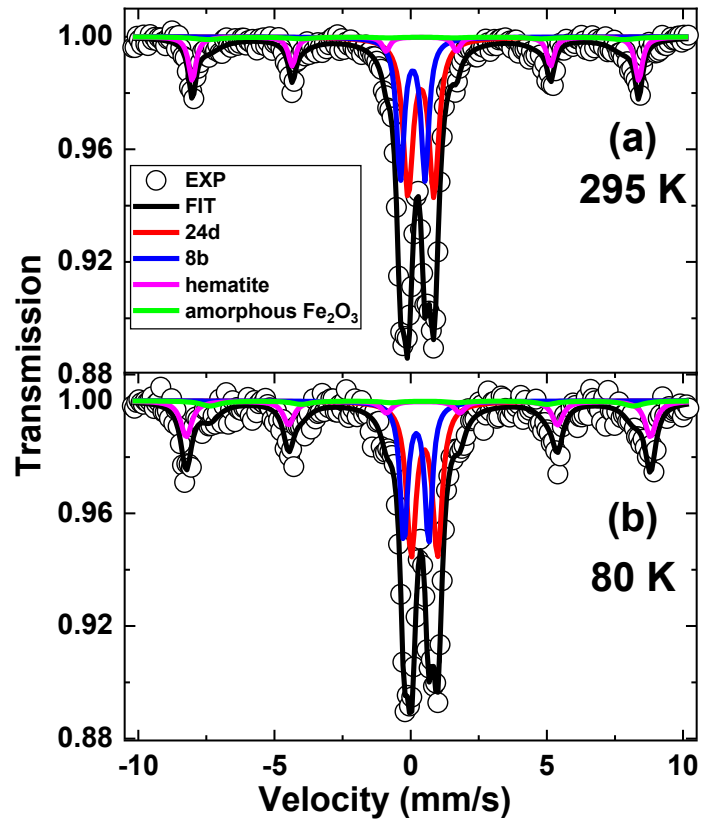

**Figure 1.** Fittings of $^{57}$Fe Mössbauer spectra of the as-synthesized sample measured at 295 K and 80 K. The open circles represent experimental data whereas the solid lines represent curve fittings. One doublet represents Fe$^{3+}$ occupying 24d sites (solid red line) and the other doublet represents Fe$^{3+}$ occupying 8b sites (solid blue line). The sextets are assigned to crystalline (solid magenta line) and amorphous (solid green line) hematite.

The Mössbauer sextet (1) is assigned to crystalline α-Fe$_2$O$_3$ in a weakly ferromagnetic-like spin state for both temperatures, as suggested by the quadrupole splitting values of $\Delta$ = -0.21 mm/s at $T$ = 295 K and $\Delta$ = -0.17 mm/s at $T$ = 80 K, typical of this magnetic ordering [34]. Therefore, no Morin transition ($T_M$) is expected in the range of 295 - 80 K. Moreover, according to Amin and Arajs [35] $T_M$ = 264.2 - 2194/$d$, where $d$ is the nanoparticle diameter (in nm) and $T_M$ is the Morin temperature (in K units). Then, using the lower temperature value ($T_M$ = 80 K), one can estimate the mean size of the as-synthesized α-Fe$_2$O$_3$ nanoparticles below 12 nm. This result agrees with the mean size estimated by XRD analysis (8 nm). According to the literature, the Morin temperature decreases as the mean particle size decreases, tending to be quenched for particles smaller than about 8 nm in mean size [35,36]. Negative values of quadrupole splitting are also found for sextet (2), indicating weakly ferromagnetic coupling for this component, at both temperatures (80 and 295 K). Bulk hematite is a weak ferromagnet below the Néel temperature (948 K ≤ $T_N$ ≤ 963 K) which undergoes a magnetic phase transition (to antiferromagnet and presenting spin reorientation) at the Morin temperature ($T_M$ ≈ 263 K). For the doublets, as can be seen in Table 1, the low temperature (80 K) isomer shift ($\delta$) and quadrupole splitting ($\Delta$) of 24d site, are 0.61 mm/s and 0.96 mm/s, respectively, which are higher than the 0.31 mm/s and 0.90 mm/s found for the 8b site. The larger values of both $\Delta$ and $\delta$, obtained for 24d site, agree well with the values reported by Nell *et al.* [37] in natural and synthetic samples of FeMnO$_3$. However, the $\Delta$ values for both sites are significantly higher than the values reported at room temperature in other studies [27,37]. This is likely related to the preparation method of the samples used here, which results in more distorted nanocrystals than those reported in the literature. The area ratio of the doublets, Area$_{24d}$/Area$_{8b}$= 1.79 was extracted at 80 K. This value will be used as a constraint for the Rietveld analysis. For the Mössbauer sextet (2), the unusual large linewidth value ($\Gamma$ = 0.87 mm s$^{-1}$, $T$ = 80 K) obtained for the amorphous hematite confirms the degree of amorphization, with the decrease of the hyperfine field, as compared to the crystalline hematite (see Table 1). Moreover, Kolk *et al.* [38] studying hematite nanoparticles have also reported the presence of an additional component with broader linewidth and hyperfine field of about 480 kG, which was attributed to amorphous α-Fe$_2$O$_3$, whereas the crystalline hematite showed hyperfine field of 525 kG at 15 K.

Table 1. List of hyperfine parameters obtained from the fits of the Mössbauer spectra, where $\delta$ is the isomer shift, $\Delta$ is the quadrupole splitting, $\Gamma$ is the linewidth, Area is the percentage area of the corresponding subspectrum, and $B_{hf}$ is the hyperfine field. $\delta$ is given relative to $\alpha$–Fe.

| T(K) | Subspectrum | $\delta$ (mm s$^{-1}$) | $\Delta$ (mm s$^{-1}$) | $\Gamma$ (mm s$^{-1}$) | Area (%) | $B_{hf}$ (kG) |
|---|---|---|---|---|---|---|
| 80 | Doublet (1), 24d | 0.61(1) | 0.96(1) | 0.42(1) | 41(2) | - |
|  | Doublet (2), 8b | 0.31(1) | 0.91(1) | 0.33(1) | 23(2) | - |
|  | Sextet (1) | 0.47(2) | -0.17(2) | 0.48(2) | 25(2) | 528(2) |
|  | Sextet (2) | 0.55(2) | -0.09(2) | 0.87(2) | 11(2) | 482(2) |
| 295 | Doublet (1), 24d | 0.49(1) | 0.95(1) | 0.43(1) | 45(2) | - |
|  | Doublet (2), 8b | 0.18(1) | 0.90(1) | 0.33(1) | 25(2) | - |
|  | Sextet (1) | 0.39(2) | -0.21(2) | 0.33(2) | 16(2) | 508(2) |
|  | Sextet (2) | 0.38(2) | -0.18(2) | 1.27(2) | 14(2) | 473(2) |

The spectral area ratio of the doublets (0.64 ± 0.02) at 80 K and the total area of all subspectra will also be used as constraint for the XRD data Rietveld refinement, as discussed later on in this report. It is worth noting that the subspectra area ratio value estimated at 295 K are slightly higher than the value estimated at 80 K. The difference in this regard is likely related to the non-negligible superparamagnetic relaxation of hematite nanoparticles (~ 8 nm) at 295 K. As expected, at room temperature, a given fraction of the nanoparticles in the sample becomes superparamagnetic (featured as doublet in the Mössbauer spectra). Therefore, it is assumed that the superparamagnetic-related subspectra area has been incorporated into the two paramagnetic doublets area (8b and 24d sites), justifying the increase of the relative area attributed to the $(Fe_{0.25}Mn_{0.75})_2O_3$ phase at room temperature. Therefore, as the relaxation is an unwanted effect, the area ratio determined at 80 K was used as the best constraint for Rietveld refinements. Moreover, this effect can also justify the strong decreasing of δ values found for the doublets at 295 K when compared with the corresponding values observed at 80 K.

## 3.2 X-ray diffraction

Figure 2 shows the room-temperature XRD pattern of the as-prepared powder sample. The XRD data was refined using the Rietveld refinement method (GSASII software). The refinement analysis indicated formation of a cubic structure, space group Ia-3, bixbyite $(Fe^{3+}Mn^{3+})O_3$ type structure as the majority phase (~86 mol%), and another minority phase identified as hematite (~14 mol%) as shown in Fig. 2. The parameters obtained from the refinement are listed in Table 2.

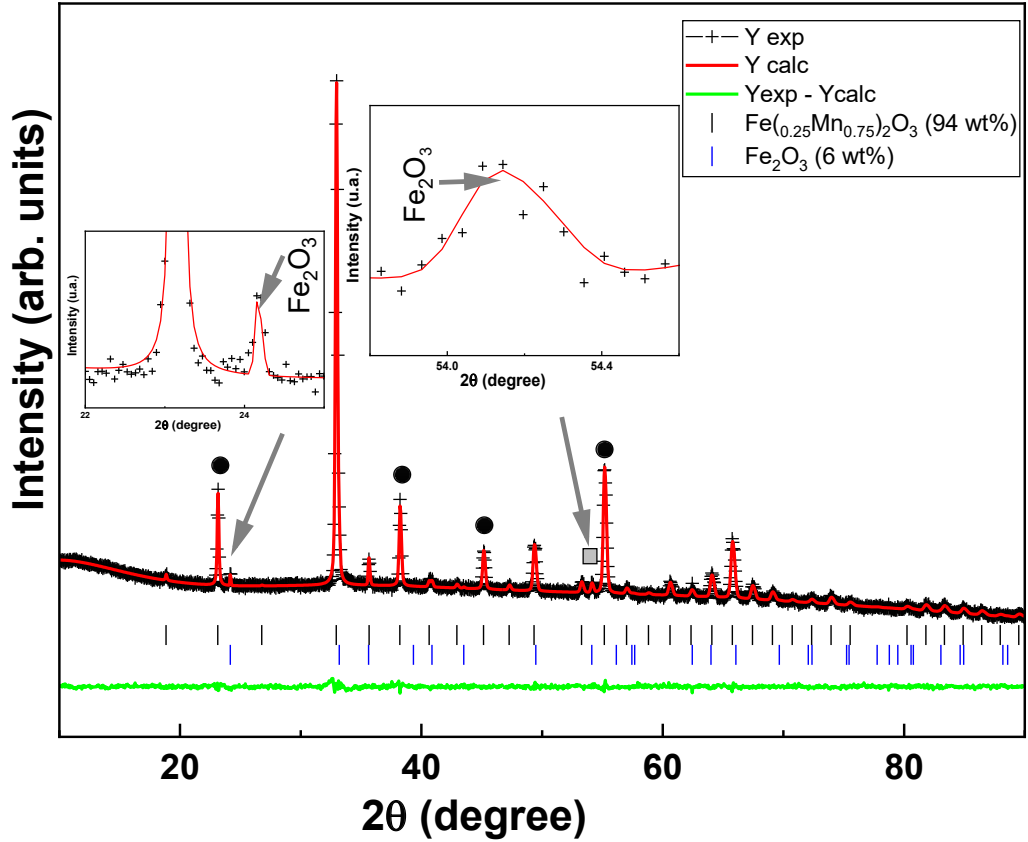

**Figure 2.** Room temperature XRD pattern of the as-synthesized sample, with the calculated data represented by the solid red line and black symbols indicating experimental data. The solid green line at the bottom shows the difference between the experimental (Yexp) and calculated (Ycalc) data. Bragg's reflections of the standard $(Fe_{0.25}Mn_{0.75})_2O_3$ and $\alpha$-$Fe_2O_3$ phases are indicated by vertical black and blue ticks, respectively.

The fractional positions for the bixbyite phase as well as the fitting parameters of the $\alpha$-$Fe_2O_3$ phase are included in Table 2. A lattice constant of $a = 9.412(1)$ Å corresponding to the bixbyite phase was found, which is consistent with the value reported in the literature $(9.41 \pm 0.03)$ Å for the $FeMnO_3$ formed in the bixbyite phase and prepared via mechano-synthesis [18].

Table 2. Structural and statistical parameters obtained from the Rietveld refinement of x-ray diffraction. $V$ is the cell volume, GOF is the goodness of fit and $\chi^2$ is the chi-square quality parameter. $\chi^2 = 1.69$; GOF = 1.30.

| | $a$ (Å) | $b$ (Å) | $c$ (Å) | $V$(Å)³ | Space group | Phase percentage (mol%) |
|---|---|---|---|---|---|---|
| $(Fe_{0.25}Mn_{0.75})_2O_3$ | 9.412(1) | 9.412(1) | 9.412(1) | 833.818(1) | Ia-3 | 86.2 |
| $\alpha$-$Fe_2O_3$ | 5.036(1) | 5.036(1) | 13.730(2) | 301.514(1) | R-3c | 13.8 |

| Fractional positions of $(Fe_{0.25}Mn_{0.75})_2O_3$ | | | | | |
|---|---|---|---|---|---|
| Atom | Wyckoff Position | $x$ | $y$ | $z$ | Occupancy |
| $Fe_1$ | 8b | 0.2500(2) | 0.2500(2) | 0.2500(2) | 0.356 |
| $Mn_1$ | 8b | 0.2500(2) | 0.2500(2) | 0.2500(2) | 0.644 |
| $Fe_2$ | 24d | 0.0308(2) | 0.0000 | 0.2500(2) | 0.212 |
| $Mn_2$ | 24d | 0.0353(2) | 0.0000 | 0.2500(2) | 0.788 |
| O | 48e | 0.3342(2) | 0.1048(2) | 0.1196(2) | 1.0000 |

| Fractional positions of $Fe_2O_3$ | | | | | |
|---|---|---|---|---|---|
| Fe | 12 | 0.0000 | 0.0000 | 0.3544(3) | 1.0000 |
| O | 18 | 0.3723(2) | 0.0000 | 0.2500(2) | 1.0000 |

The stoichiometry $(Fe_{0.25}Mn_{0.75})_2O_3$ was deduced from the Fe and Mn occupancies (see Table 2). In agreement with the results obtained from MS, it was verified that the Fe atoms site occupancy ratio (24d/8b) is 1.79. It was also possible to calculate the ratio of Fe atoms in the $(Fe_{0.25}Mn_{0.75})_2O_3$ phase with respect the total amount of Fe atoms (all phases), resulting in (0.61 ± 0.02). This finding is also in agreement with the result obtained from MS, considering the uncertainties, and allows one to conclude that this is the most reliable stoichiometry for the as-synthesized sample.

The mean crystallite size ($<L_{XRD}>$) of the $(Fe_{0.25}Mn_{0.75})_2O_3$ phase was estimated using the modified Scherrer's equation. This was accomplished while plotting $\ln\delta$ versus $\ln(1/\cos\theta)$ to obtain the y-axis intercept $\ln(k\lambda/<L_{XRD}>)$, using least square linear regression, where $k$ is a constant (0.89 for spherical nanoparticle), $\lambda$ is the wavelength of the x-ray (Cu-$K\alpha$), $\delta$ is the full width at half maximum (FWHM) of the x-ray diffraction line, and $\theta$ is the corresponding diffraction angle [39]. The mean crystallite size $<L_{XRD}>$ ~ 48 nm was estimated for the FeMnO$_3$ phase using the (211), (400), (332) and (440) Bragg planes (see solid black circles in Fig. 2). To estimate the mean crystallite size of the secondary $\alpha$-Fe$_2$O$_3$ phase, the Scherrer's equation was used only for the (116) plane (see the square gray in Fig. 2), once it represents the highest XRD intensity peak observed for this phase. Then, $<L_{XRD}>$ ~ 8.0 nm was estimated for the $\alpha$-Fe$_2$O$_3$ phase.

## 3.3 Morphology

Figure 3a shows a representative TEM micrograph of the FeMnO$_3$ like sample produced at 900 °C. As see in Figs. 3a and 3b, irregularly shaped particles with a wide range of sizes extending from 20 to 820 nm are observed. The crystal structure of the (Fe$_{0.25}$Mn$_{0.75}$)$_2$O$_3$ was also examined by selected-area electron diffraction (SAED). The selected area electron diffraction (SAED) pattern, obtained from the (Fe$_{0.25}$Mn$_{0.75}$)$_2$O$_3$ sample (highlighted by the white circle in Fig. 3b), reveals a distinctive arrangement of broad concentric diffraction rings. Notably, the rings correspond to crystallographic planes such as (200), (211), (321), (222), (411), and (422), attributing them to the cubic phase of the (Fe$_{0.25}$Mn$_{0.75}$)$_2$O$_3$ sample (depicted in Fig. 3d). Remarkably, the absence of the hematite phase in this SAED pattern suggests that it does not coat the particle surface. Instead, a secondary phase is formed. This intriguing observation is underscored by the clear absence of the aforementioned electron diffraction associated with hematite in this particular region. The crystallographic planes of each diffraction ring are explicitly indicated in Fig. 3d.

The EDX spectrum recorded from the as-synthesized sample is shown in Fig. 3e. As observed, the peaks corresponding to Mn, Fe and O confirm the presence of these elements in the major phase (Fe$_{0.25}$Mn$_{0.75}$)$_2$O$_3$. The C and Cu signal appearing in the EDX spectrum is due to the tape and grid used for sample preparation, respectively.

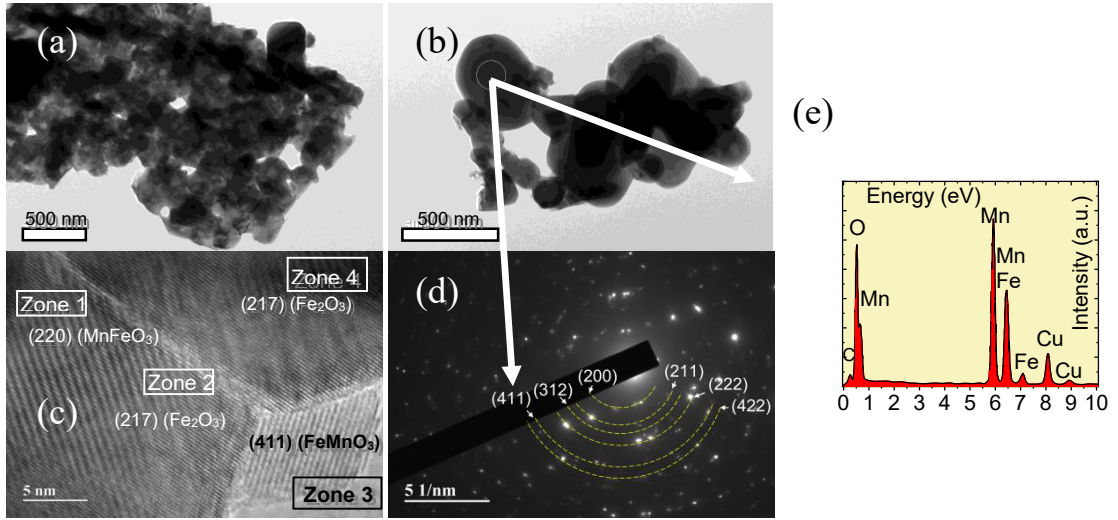

**Figure 3.** (a) and (b) TEM images of the as-synthesized sample (scale bar in nm). (c) HRTEM image of the as-synthesized sample with Zone 1 and Zone 3 referring to interplanar distance of the FeMnO$_3$ phase, whereas Zone 2 and Zone 4 refer to the α-Fe$_2$O$_3$ phase (scale bar in nm). (d) SAED pattern for the as-synthesized sample (scale bar in nm). (e) Shows the EDX spectrum of the as-synthesized sample recorded in the indicated position (white circle) of panel (b).

## 3.5 Raman Spectroscopy

The crystallographic symmetry is the bixbyite structural type with the *Ia-3* space group having 22 Raman active modes described by: $4A_g + 4E_g + 14F_g$ [40], 10 inactive modes $5A_u + 5E_u$, and 16 $T_u$ IR modes [41]. Despite the large number of Raman active modes (22), the number of modes actually observed in the Raman spectrum is reduced [40]. Figure 4 shows the room-temperature Raman spectrum of the as-synthesized powder sample in the 200-800 cm$^{-1}$ range. According to the above-mentioned crystallographic symmetry, the bixbyite structural type is the source of the Raman active modes in the spectrum, which is consistent with the XRD result. Furthermore, the bands exhibit broadness, which is unexpected given the higher average particle size reported by XRD and TEM. Concurrently, this implies a significant degree of disorder, particularly at the particle surface, as the Raman shows a greater surface contribution than the XRD.

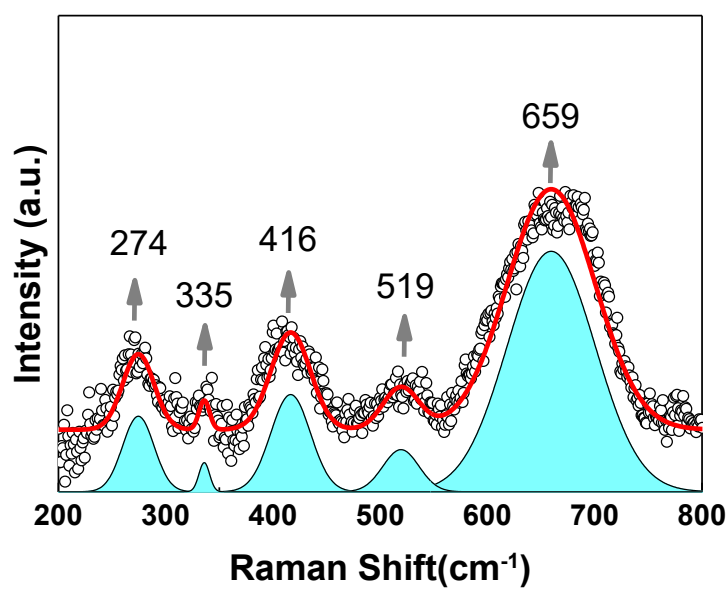

**Figure.4.** Raman spectrum of the as-synthesized sample. The experimental data are represented by open black symbols whereas the solid red line portrays the best fitting obtained by incorporating blue colored components.

The characteristic vibrational bands observed at 659 cm$^{-1}$, 519 cm$^{-1}$ and 416 cm$^{-1}$ correspond to three of 14 $F_g$ modes which are labeled as $F_g^{(1)}$, $F_g^{(2)}$, and, $F_g^{(3)}$, respectively. The band located at 335 cm$^{-1}$ is assigned to the $E_g + F_g$ mode, meanwhile, the 274 cm$^{-1}$ band is linked to the $E_g$ mode. To display the effect of the presence of iron in the bixbyite holding matrix we can highlight the spectrum reported by Chen *et al.* which displays the $F_g^{(1)}$ located at 652 cm$^{-1}$ linked to the Mn$^{3+}$-O mode vibration. When iron is present in the Mn/FeO$_6$ octahedral, as in the case of the FeMnO$_3$, vibrations can cause the Raman frequencies to shift to lower values; these downshift being also observed in the $F_g^{(2)}$ and $F_g^{(3)}$ vibrational modes [42].

## 3.5 Surface composition analysis

XPS measurements were performed to examine the oxidation state of Fe and Mn and the presence of oxygen specimens on the surface of the nanoparticles. The background was modeled using a Shirley-type function implemented in Casa XPS software, and all positions were corrected using the expected 284.6 eV position of the C 1s binding energy. The high-resolution spectrum of Fe 2p, displayed in Fig. 5a, clearly shows Fe 2p$_{1/2}$ and Fe 2p$_{3/2}$ enlarged photoelectron lines (PL), suggesting that there are two sets of peaks corresponding to two different oxidation states. In this regard, a good fit was obtained using two PL for each Fe 2p peak, deconvoluted into two mixed Gaussian-Lorentzian-shaped peaks. The doublets binding energies (BEs) obtained have been located at 710.2 eV/723.6 eV, and 712.5 eV/725.9 eV, which were assigned to $Fe^{2+}$ and $Fe^{3+}$ oxidation states, respectively. In addition, satellite peaks were observed at 715.1 eV/719.7 eV and 728.3 eV/733.3 eV which are also characteristic of $Fe^{2+}$ and $Fe^{3+}$, respectively. The percentage of each species was estimated from the integrated areas of each peak to be around ~43% of $Fe^{3+}$ and ~57 % of $Fe^{2+}$. These values are consistent with the crystalline state identified with different occupational sites for Fe atoms using XRD and MS.

As shown in Fig 5b, the fitting of the Mn 2p core level high-resolution XPS spectra reveals two component features as well, which are centered at 641.0eV/643.1eV and 652.4 eV/654.2 eV. The Mn $2p_{3/2}$ and Mn $2p_{1/2}$ features centered respectively at binding energies ~641.0eV and ~ 652.4 eV, are indication of the presence of $Mn^{3+}$ ions with a spin-orbit split equals to ~11.4 eV. Moreover, the binding energies ~643.1 eV and 654.2 eV are indication of presence of presence of $Mn^{4+}$ ions with a spin-orbit split equals to ~11.1 eV. The satellites of the Mn $2p_{1/2}$ peak are located at 663.3 eV and 667.4 eV, with a separation between them of about 9 eV (see Fig. 5b), which is a fingerprint of the $Mn^{3+}$ and $Mn^{4+}$ oxidized state, respectively [43].

The XPS spectrum of O 1s, revealing three peaks, with binding energies of approximately 529.9, 531.3 and 533.0 eV, present integrated areas of 34%, 45% and 20%, respectively. Similar observation is reported in $BaTiO_3$ [44], which has been assigned to $O^{2-}$ ions, $O^{1-}$ ions, and $O^{Chem}$ species. Regarding the high percentage of $O^{Chem}$ species observed (20%) it could be associated with oxygen vacancies present on the surface of the $FeMnO_3$ nanocrystals. It is worth noting that this technique can detect only oxides at the surface (up to 2 nm deep into the nanoparticle). The core of the nanoparticles may have different composition, very likely prevailing the oxide with $Fe^{3+}$ and $Mn^{3+}$.

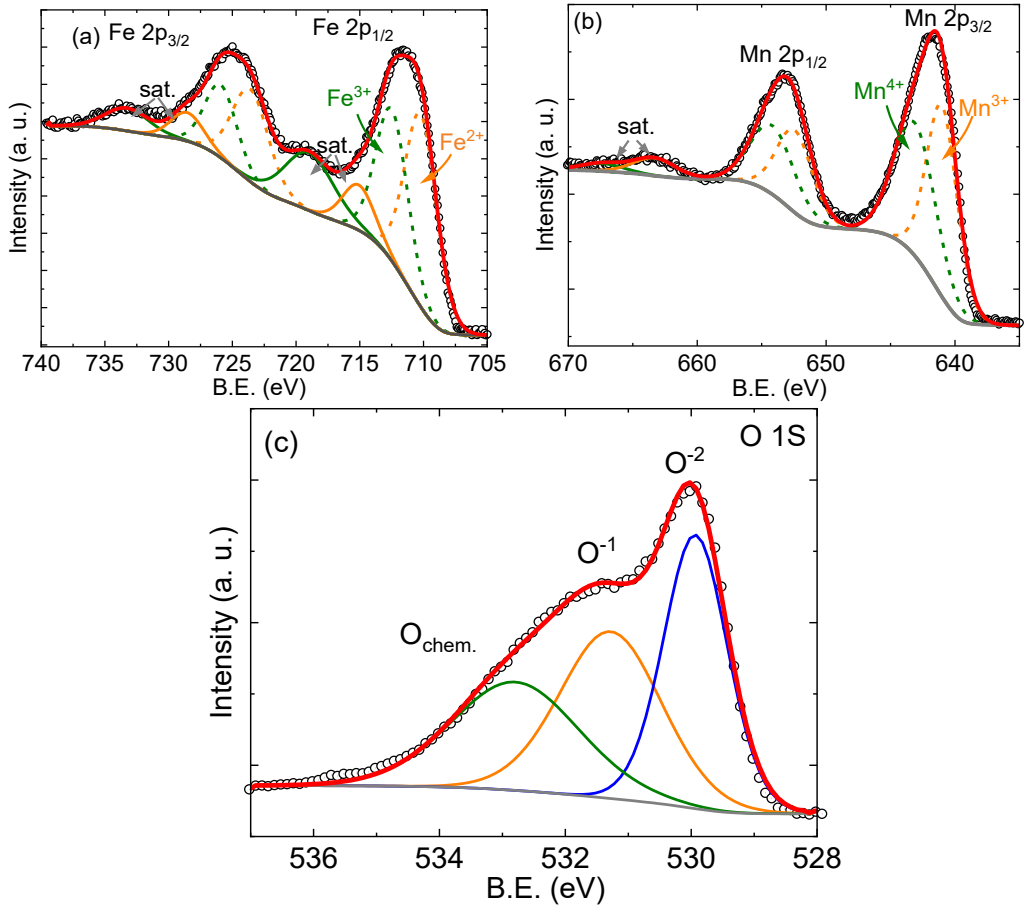

**Figure 5.** XPS spectra of the as-synthesized sample, with panels (a), (b), and (c) representing the high-resolution XPS spectra of Fe 2p, Mn 2p, and O 1s, respectively.

## 3.6 Magnetic measurements

Figures 6a and 6b show the magnetization hysteresis loop for the as-synthesized sample at 300 K and 5 K, respectively. The magnetization at 300 K is consistent with the presence of mainly a paramagnetic behavior, superimposed to superparamagnetic response, the latter saturating at relatively small magnetic field. The dashed line in Fig. 6a indicates the saturation magnetization of the superparamagnetic contribution credited to the $\alpha$-$Fe_2O_3$ secondary phase [45,46] according to the percentage of phases listed in Table 2. The inset in Fig. 6a shows the central part of the 300 K $M$ vs $H$ curve (magnetic field up to $\pm$ 10 kOe). Therefore, the magnetization data confirm the presence of $\alpha$-$Fe_2O_3$ secondary phase as determined from the Rietveld refinement data analysis. Moreover, the negligible value for the coercive field at 300 K confirms the superparamagnetic regime for these phases. It is worth noting that the particles of the hematite phase have been determined as blocked at 300 K, once the Mössbauer spectra for them are two magnetic sextets.

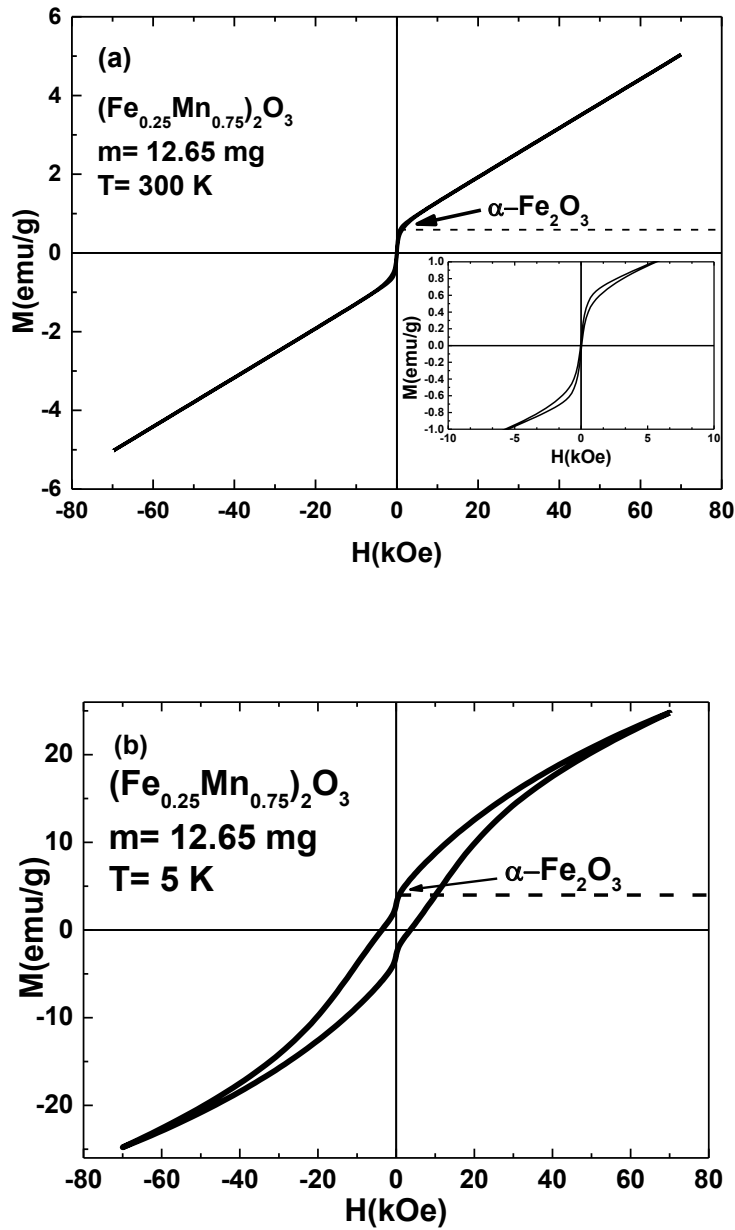

**Figure 6.** Magnetization hysteresis loops in the ± 70 kOe range obtained (a) at 300 K (the inset shows a zooming of the 300 K plot in the ± 10 kOe range) and (b) at 5 K (note the coercivity and magnetic hysteresis in the low field range). Mass of the measured powder is quoted in the legend and the dashed line in panel (a) represents mainly the contribution of the $\alpha$-$Fe_2O_3$ (secondary) considering the phase percentage obtained by the Rietveld refining (see Table 2) and the saturation magnetization of bulk $\alpha$-$Fe_2O_3$ at 300 K ($Ms = 84$ emu/cm$^3$).

This disagreement is due to different time of measurements involved for the magnetization ($10^2$ s) and Mössbauer spectroscopy ($10^{-8}$ s) techniques, resulting in higher $T_B$ for the latter (MS). While comparing Figs. 6a and 6b, one finds remarkable differences between the *M* vs *H* curves collected at 300 K and 5 K, reflecting the magnetic evolution of the dominant phase. For instance, the coercive field increase, which is discussed in detail in the next paragraphs.

The temperature dependence of zero-field-cooled (ZFC) and field-cooled (FC) magnetization curves of the as-synthesized sample obtained with a magnetic field of 20 Oe is shown in Fig. 7. The ZFC curve presents a sharp peak at $T_C \sim 33$ K (assigned to the ferrimagnetic transition), following a sharp drop with decreasing *T*. Simultaneously, a rising feature is observed in the FC curve while decreasing *T*, showing a strong irreversibility between ZFC and FC curves. This strong irreversibility can be assigned to the presence of the secondary $\alpha$-$Fe_2O_3$ phase, besides the majority phase $(Fe_{0.25}Mn_{0.75})_2O_3$. Moreover, the temperature related to the peak observed in the ZFC curve ($T_C \sim 33$ K) is lower than the value reported for the $FeMnO_3$ system prepared by mechano-synthesis ($T_C \sim 40$ K) [15,26]. This is mainly due to samples with different stoichiometry.

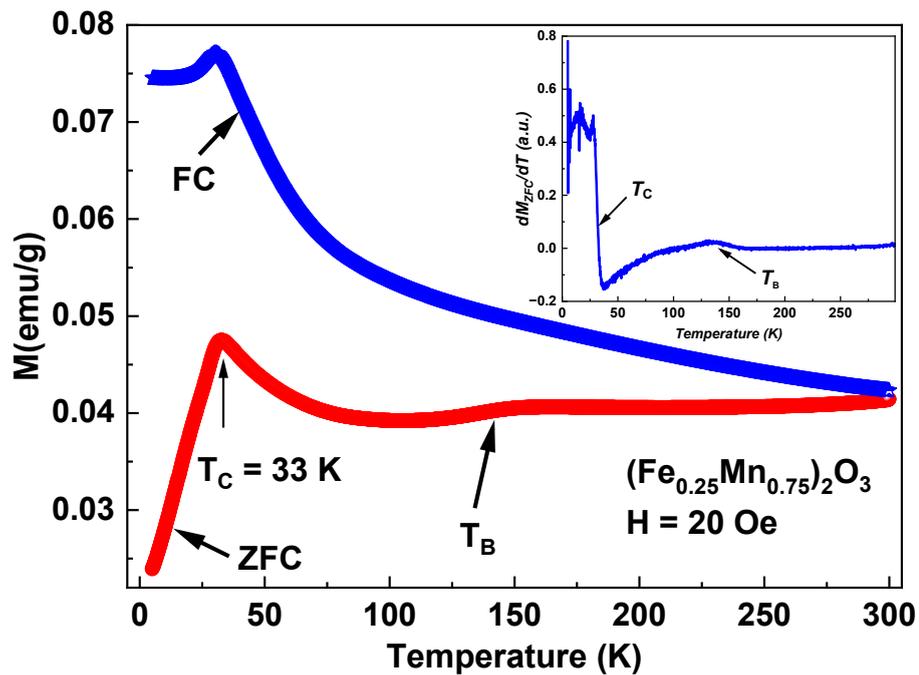

**Figure 7.** Temperature dependence of zero-field-cooled (ZFC) and field-cooled (FC) traces obtained while applying a magnetic field of 20 Oe. The inset shows the derivate ($dM_{ZFC}/dT$) used to identify the blocked temperature ($T_B$) of the $\alpha$-$Fe_2O_3$ and the ferrimagnetic transition temperature ($T_C$).

Importantly, in Fig. 7 one can observe a valley around 150 K. This feature is attributed to the blocking temperature ($T_B$) of the hematite phase. To determine the exact blocked temperature d(ZFC)/d$T$ is plotted as a function of temperature (see inset in Fig. 7). The d(ZFC)/d$T$ versus $T$ plot shows a visible downturn around 140 K, which is attributed to the blocked temperature of the $\alpha$-Fe$_2$O$_3$ phase. To determine the type of magnetic transition involved in the as-synthesized sample, the temperature dependence of the inverse magnetic susceptibility ($\chi^{-1}$) measured at 20 Oe is plotted in Fig. 8.

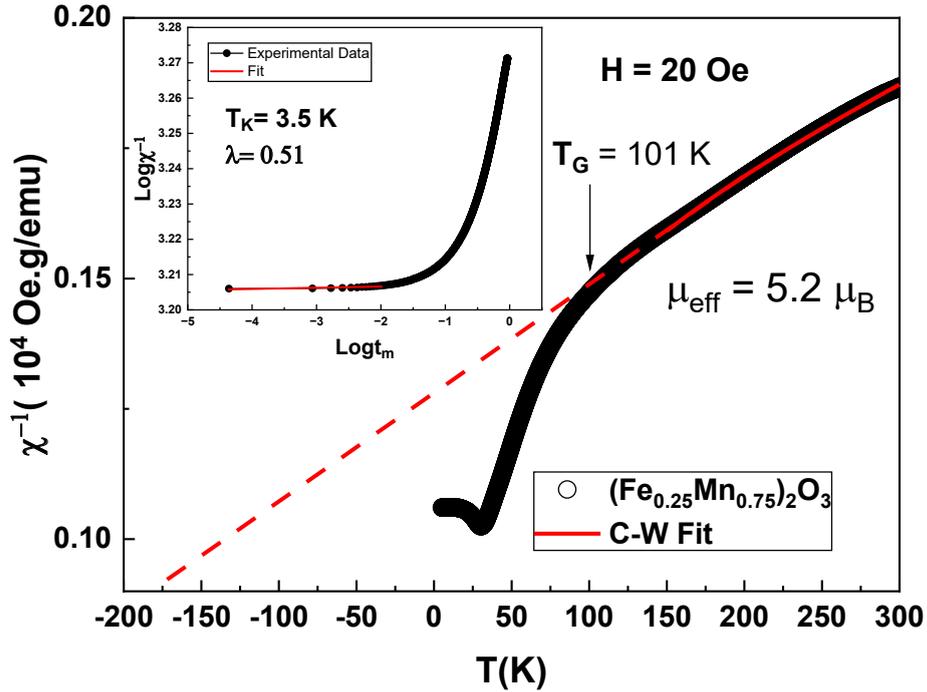

**Figure 8.** Temperature dependence of the inverse susceptibility ($\chi^{-1}$), at 20 Oe (black symbols), along with the Curie-Weiss fit (solid red line) in the 150-300 K range using Eq. 1. The inset shows the temperature dependence of the dc susceptibility (following Eq. 2) plotted in a double logarithmic scale.

Data in Fig. 8, regarding the high-temperature range (150-300 K), were fitted (solid red line) to the Curie-Weiss law [47]:

$$\chi = \frac{C}{T-\theta} \quad (1)$$

where $C=N\mu_{eff}^2/3k_B$ is the Curie constant ($N$ is the number of magnetic ions, $\mu_{eff}$ is the effective magnetic moment per ion, $\theta$ is the characteristic Curie-Weiss temperature, and $k_B$ is the Boltzmann constant.

It is worth mentioning that the Curie-Weiss data analysis (see Fig. 8) must reflect mainly the $(Fe_{0.25}Mn_{0.75})_2O_3$ phase. The best fit provides $\theta = (-177 \pm 2)$ K. The negative and relatively high $\theta$ value suggests strong AFM interactions between $Fe^{3+}$ and $Mn^{3+}$ magnetic moments in the bixbyite crystallites formed by several unit cells [48]. Moreover, they also indicate the presence of magnetic frustration in the nanocrystals [18]. The literature on $FeMnO_3$ [49] has established the frustration ratio parameter (defined as the ratio between the absolute value of the Curie-Weiss temperature, $|\theta|$, and $T_C$, i. e. $|\theta|/T_C$) as an indicator of the degree of magnetic frustration in the sample. For our $(Fe_{0.25}Mn_{0.75})_2O_3$ sample, this ratio is $|\theta|/T_C = 177/33 = 5.4$, a value smaller than the one ($|\theta|/T_C = 336/32 = 10.5$) obtained by Roth et al. for $(Fe_{0.56}Mn_{0.44})_2O_3$ [49,50], where a long-range magnetic order was not observed at low temperatures, indicating a prevailing spin glass state. However, our frustration ratio value is higher than the one ($|\theta|/T_C = 69/36 = 1.9$) obtained by S. Rayaprol et al. [18,19] for $(Fe_{0.5}Mn_{0.5})_2O_3$, in which an antiferromagnetic order is established at low temperature.

A Curie constant of $C = 0.36$ emu/mol×K was obtained from the fit. From this value and assuming the structure has approximately 16 pairs of $Fe_{0.5}Mn_{1.5}$ at sites 8b and 24d per unit cell, the experimental value of the effective magnetic moment per pair of $Fe_{0.5}Mn_{1.5}$ is $\mu_{eff}^{exp} = 5.2$ $\mu_B$. This value is higher than the value of 2.8 $\mu_B$ determined by Seifu et al. in $FeMnO_3$ compounds produced by the mechanical alloying technique [26]. This is attributed to differences in stoichiometry between the investigated samples Assuming that the electronic configuration of the system is $(Fe^{3+}_{0.5})(Mn^{3+}_{1.5})(O_3^{2-})$ and taking into account that the theoretical spin-only values are: $\mu_{Fe}^{3+} = 5.90$ $\mu_B$ and $\mu_{Mn}^{3+} = 4.90$ $\mu_B$ ($Fe^{3+}$ and $Mn^{3+}$ ions in high spin state), the theoretical value of the effective magnetic moment is estimated as [51]: $\mu_{eff}^{theo} = [(1-f) \times (\mu_{Mn}^{3+})^2 + f \times (\mu_{Fe}^{3+})^2]^{1/2} = 5.2$ $\mu_B$, where $f = 0.25$ is the percentage of Fe over the total magnetic ions of the system. This result is in excellent agreement with the experimental one, confirming that $Fe^{3+}$ and $Mn^{3+}$ ions are in the high spin state configuration.

The downward deviation of the temperature-dependent inverse magnetic susceptibility ($\chi^{-1}$) from the ideal Curie-Weiss law (see Fig. 8) signals the presence of the Griffiths phase (GP) in the system. The temperature at which $\chi^{-1}(T)$ deviates from the ideal Curie-Weiss law is known as the Griffiths temperature ($T_G$) herein ascribed to 101 K, as shown in Fig. 8. This feature is a unique characteristic of the GP.

A similar downturn was also observed in the antiferromagnetic $TbFe_{0.5}Cr_{0.5}O_3$ and $FeMnO_3$ phases fabricated by the mechano-synthesis method [52,53], in which the FM correlation among neighboring short-range FM cluster accounts for the observed deviation. Herein, $T_G$ is defined as the onset temperature, where $\chi^{-1}(T)$ data deviates from the Curie-Weiss law, and a local AFM ordered area begins to develop [54]. The GP regime is usually characterized by the temperature dependence of the inverse susceptibility, which follows a power law:

$$\chi^{-1}(T) \propto (T - T_K)^{1-\lambda} \qquad (2)$$

with $0 \leq \lambda < 1$. The power law in the previous equation is a generalized Curie-Weiss law. Here, $T_K$ is the critical temperature of the ferromagnetic (or ferrimagnetic) clusters [18] that can be estimated from $\lambda = 0$ in the Curie-Weiss regime [55,56], which is equivalent to the Curie-Weiss temperature ($\theta$). The parameter $\lambda$ appearing in the exponent shows the strength of the GP. The double logarithmic plot of the dc susceptibility against reduced temperature ($t_m = \frac{T}{T_K} - 1$), reproduced in the inset of Fig. 8, shows a linear behavior at low $t_m$ values, and confirm the proposed GP. The fitted value of $\lambda$ is 0.51 for the as-synthesized $(Fe_{0.25}Mn_{0.75})_2O_3$ sample which is in the range ($0 \leq \lambda < 1$) expected from a system exhibiting the Griffiths phase. The solid red line in the inset of Fig. 8 represents the best fit of the $\chi^{-1}$ versus $T$ data using Eq. 2, in the range of 5K $< T < T_G$. Actually, Eq. 2 represents the temperature dependence of the order parameter ($\chi^{-1}$) in the context of the classical Landau second-order phase transition, with the critical exponent ($\delta = 1-\lambda$) very much close to ½ [57].

### 3.7 Magnetic resonance

The MR spectra obtained from the as-synthesized sample at a fixed frequency of 9.50 GHz, in the temperature range of 4.3 - 293 K are presented in Fig. 9. Throughout this temperature range, a strong and broad resonance signal, with a nearly symmetrical shape, can be observed. This shape is strictly Lorentzian, suggesting a strong interaction between metal ions through the exchange interaction [58]. Moreover, it suggests that the majority $(Fe_{0.25}Mn_{0.75})_2O_3$ phase dominate the spectra. By utilizing $g = h\nu/\mu_B H_R$, where $\nu$ denotes the spectrometer's operating frequency (9.50 GHz), $\mu_B$ represents the Bohr magneton and $H_R$ represents the extracted MR field, the temperature dependence of the extracted $g$-values can be plotted, as shown in the inset of Fig. 9. The $g$-values for the as-synthesized ($Fe_{0.25}Mn_{0.75})_2O_3$ nanocrystals have been found to decrease systematically with increasing temperature, in agreement with the behavior found to perovskites compounds [59,60]. This behavior is largely attributed to a decreasing of the magnetic moments occurring as the temperature rises [18,61]. Specifically, these values decreased from (2.93 ± 0.01) at 50 K to (2.18 ± 0.01) at 300 K.

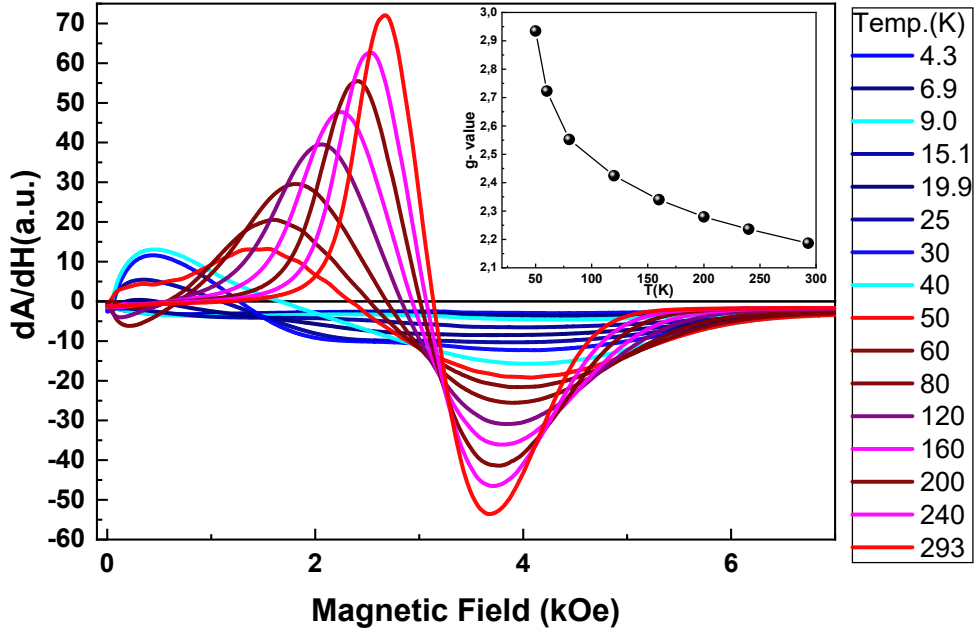

**Figure 9.** X-band MR spectra, representing the first derivate of the absorption, recorded from the as-synthesized sample. The spectra were recorded at various temperatures, ranging from 4.3 K to 293 K. The temperature dependence of the $g$- factor is also plotted in the inset (the solid line is just to guide the eyes).

The inset of Fig. 10 shows $H_R$ vs $T$, where the vertical dotted black lines indicate $T_C$, $2T_C$, and $T_G$ temperatures. For $T_G < T < 300K$, the majority $(Fe_{0.25}Mn_{0.75})_2O_3$ phase is paramagnetic and $H_R$ reduces smoothly with decreasing $T$, due to the enhancement of the magnetic moment [18,61]. For $2T_C < T < T_G$, the $(Fe_{0.25}Mn_{0.75})_2O_3$ phase is still in the paramagnetic state but with the presence of clusters formed at $T_G$, and $H_R$ decreases sharply due to the enhancement of the magnetic moment and the presence of these clusters. Importantly, the Griffiths-like phase is an intermediate state between the disordered paramagnetic and the ordered ferromagnetic (or ferrimagnetic) state, where it begins to appear magnetic interaction between the FM (or ferrimagnetic) clusters. In the study of FeMnO$_3$ nanoparticles prepared by mechano-synthesis method, Rayaprol *et al.* [18] showed that at $T\sim2T_C$ a change in the unit volume cell occurred, and they argued that there is a strong coupling between the structure and the magnetic ordering occurring at $T_C$, provoking large variations of magnetic entropy in this temperature range. Although another method of synthesis has been herein employed, $H_R$ for $T < 2T_C$ also reduces sharply while decreasing the temperature, confirming that the as-synthesized sample can also show similar behavior on $H_R$ from 66 K ($2T_C$) down to 4.2 K. Therefore, this sharp drop is attributed to the increasing magnetic interaction influenced by structural changes and long-range ferrimagnetic coupling as the system approaches $T_C$, as well as the spin-glass-like behavior occurring at the surface of the nanoparticles. Figure 10 shows the variation of peak-to-peak line width ($\Delta H_{pp}$) as function of temperature for the (Fe$_{0.25}$Mn$_{0.75}$)$_2$O$_3$ nanocrystals, showing a sharp peak at $T_{max}$ = 42 K (~1.27 $T_C$), where $T_{max}$ is assigned to the temperature where the $\Delta H_{pp}$ is maximum. It is known that for systems showing spin-glass-like behavior, $\Delta H_{pp}$ presents noticeable broadening below the spin-glass temperature [62]. As shown in Fig. 10, higher value for $T_{max}$ (43 K) while compared to $T_C$ (33 K) is attributed to a surface spin-glass-like behavior of (Fe$_{0.25}$Mn$_{0.75}$)$_2$O$_3$ nanocrystals.

Actually, the study of spin-glass-like systems has relied on the temperature dependence of $\Delta H$pp to obtain information about the spin freezing phenomenon and the corresponding freezing temperature ($T_f$). Recently, an exponential relationship between $\Delta H_{pp}$ and $T$ for diluted magnetic semiconductors exhibiting spin-glass behavior has been proposed [63]:

$$\Delta H_{pp} = \Delta H_\infty + \Delta H_0 (1 - \theta/T) e^{-(T_f/T)} \tag{3}$$

The above-presented relationship incorporates parameters such as $\Delta H_0$ and $\Delta H_\infty$, which describes $\Delta H_{pp}$ respectively at low and high temperatures, and correlates with the concentration of magnetic ions and the dominant magnetic interaction above the transition temperature. The Curie-Weiss temperature ($\theta$) is included in the pre-exponential term, whereas the exponential term allows for determination of the freezing temperature ($T_f$). Black symbols in Fig. 10 represent the temperature dependence of $\Delta H_{pp}$, with the experimental data fitted to Eq. 3 in the low temperature region, up to 50 K. While Eq. 3 successfully fits the $\Delta H_{pp}$ versus $T$ data below 50 K, it fails to account for the experimental data above this temperature. Notably, the $\Delta H_{pp}$ values systematically decrease from about 4200 Oe to about 3600 Oe as the temperature is increased from about 50 K to 300 K, representing the opposite trend observed below about 50 K. Importantly, below and above this temperature (50 K), the concavity is upward and downward respectively, signaling a remarkable change in the dominant magnetic behavior.

A key issue in the present study was to analyze the $\Delta H_{pp}$ versus $T$ data in the whole temperature range (4.3 K to 293 K). In order to accomplish this goal, an additional term $\Delta H_{spm} \tanh(T_{spm}/T)$ was included into Eq. 3 [64]:

$$\Delta H_{pp} = \Delta H_\infty + \Delta H_0(1 - \theta/T)e^{-(T_f/T)} + \Delta H_{spm} \tanh(T_{spm}/T) \qquad (4)$$

This extra term has proven to be successful in describing the $\Delta H$pp of superparamagnetic particles in a wide range of temperature. Actually, it is related to thermally-induced jumps between two energy minima, which correspond to two distinct orientations of nanoparticle's magnetic moment with respect to the easy axis of magnetization [65,66]. The third term (extra term) on right-hand side of Eq. 4 includes the pre-factor $\Delta H_{spm} = 5g\beta Sn/R^3$, with $S$, $n$ and $R$ representing the effective spin of the magnetic center, number of magnetic centers inside the superparamagnetic particle and average particle-particle distance, respectively. Also, included into the extra term is a characteristic temperature $T_{spm} = \Delta E/2k_B$, where $\Delta E$ is the energy barrier between the two orientations of the nanoparticle's magnetic moment. The solid red line in Fig. 10 represents the best fit of the experimental data while using Eq. 4, with the following fitted values: $\Delta H_\infty = (358 \pm 13)$ Oe, $\Delta H_0 = (84 \pm 2)$ Oe, $\theta = (-150 \pm 5)$ K, $T_f = (5.1 \pm 0.1)$ K, $\Delta H_{spm} = (3849 \pm 10)$ Oe, and $T_{spm} = (1194 \pm 5)$ K. Some of these values can be compared with the values obtained experimentally, as instance $\theta = -177$ K (see legend of Fig. 8) and $948$ K $\leq T_N \leq 963$ K obtained for bulk $\alpha$-Fe$_2$O$_3$ [67]. Higher value obtained for $T_{spm}$ should be related to the nanosized characteristic of $\alpha$-Fe$_2$O$_3$ particles.

Importantly, it is herein claimed that the extra term included into Eq. 4 describes mainly the contribution of the secondary phase, namely the superparamagnetic hematite (estimated XRD mean size of 8 nm), being dominant at high temperatures (above 50 K), although contributing to a relatively small change in the $\Delta H_{pp}$ values, roughly from 3600 Oe to 4200 Oe. Broadening of the MR line while lowering the temperature of superparamagnetic hematite has been widely reported in the literature [68-70]. The successful fitting of the data in the full temperature range of investigation has allowed for a better understanding of the model picture. Overall, the findings have provided valuable insights into the behavior of magnetic nanoparticles at varying temperatures.

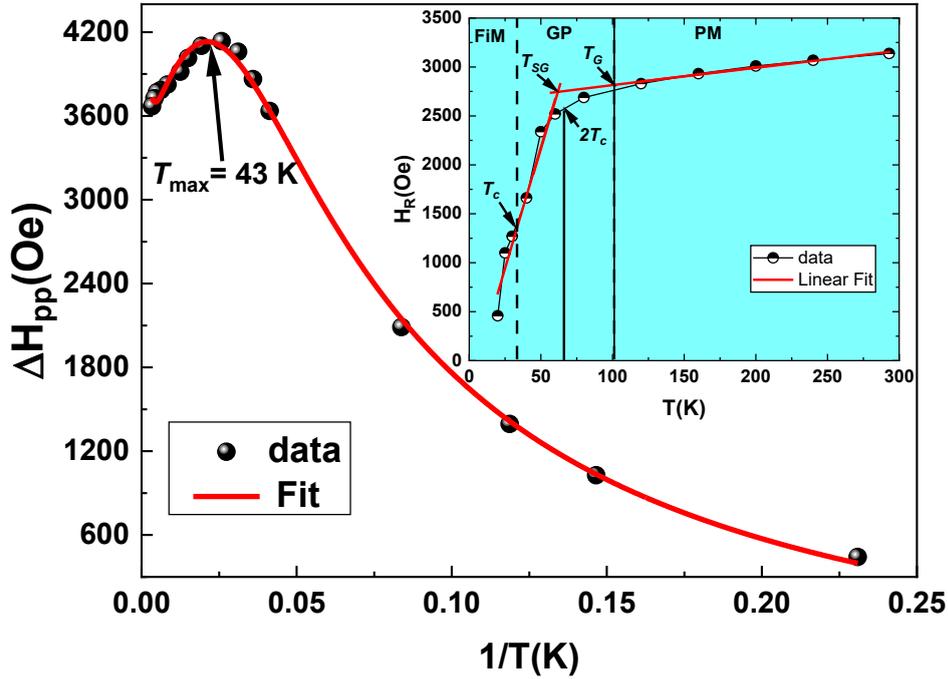

**Figure 10.** Experimental values of the magnetic resonance linewidth ($\Delta H_{pp}$) tracked as a function of temperature ($T$), as indicated by the solid black symbols. The solid red line represents the best fit using Eq. 5. The inset shows the temperature dependence of the resonance field (solid black-white symbols).

Indeed, the present report emphasizes the applicability of the MR technique, while using the $\Delta H_{pp}$ versus $T$ data, to unveil the magnetic contributions of a multi-phase nanomaterial presenting distinct magnetic ordering. As indicated by both Mössbauer and XRD data evaluation, the as-synthesized sample comprises mainly two distinct nanophases; the majority (~86 mol%) iron manganese trioxide $(Fe_{0.25}Mn_{0.75})_2O_3$ phase, with mean size around 48 nm, and the minority (~14 mol%) hematite ($\alpha$-$Fe_2O_3$) phase with mean size around 8 nm. Typically, different iron oxide nanophases (e.g. iron sulfate, iron oxide, Cd/Zn/Cu/Ni/Mn-ferrite), pristine or surface-dressed, synthetic or extracted from living organisms, present broad MR lines (up to about 4 kOe) and resonance field below about 3.5 kOe [71,72,64,65,68,73,74]. Therefore, it is not surprising that the majority $(Fe_{0.25}Mn_{0.75})_2O_3$ phase dominates the MR spectra shape in the wide temperature range of our investigation (4.3 - 293 K). Nevertheless, the signature of the minority hematite phase is clearly identified in the temperature dependence of $\Delta H_{pp}$ above about 50 K, as described by Eq. 4. Whereas the larger (48 nm) magnetic phase (iron manganese trioxide) dominates the $\Delta H_{pp}$ trend in the lower temperature range (below about 50 K) the smaller (8 nm) magnetic phase (hematite) contributes remarkably to the $\Delta H_{pp}$ trend in the higher temperature range (above about 50 K). It is worth noting that the spin-glass-like behavior should be concentrated at the surface of the nanocrystals (~0.6 nm tick), whereas the core can be magnetic, as reported in other magnetic nanoparticles [60,75]. Moreover, it is not completely ruled out that the surface spin-glass-like behavior can also occur for $\alpha$-Fe$_2$O$_3$ nanoparticles. The third term on the right hand-side of Eq. 4 accounts for the magnetic behavior of the hematite phase, with negligible increment below 60 K, but with non-negligible contribution above 60 K. Therefore, it is not surprising that Eq. 3 can fit nicely the $\Delta H_{pp}$ versus $T$ data below about 60 K and fails to perform the fitting in the whole temperature range (4.3-293 K). Likewise, it is not surprising the need of an extra term to account for the superparamagnetic contribution of the hematite phase above 60 K, as included into Eq. 4. Moreover, as the temperature drops below 60 K, it is claimed that a phase transition occurs from a continuous paramagnetic phase to a spin-glass-like phase in the shell layer of the (Fe$_{0.25}$Mn$_{0.75}$)$_2$O$_3$ phase. Regarding a rough estimation of the start of this transition, two linear fits have been carried out for $H_R$ vs $T$, one of them from 20 up to 50 K and the other from 120 to 295 K, where the intersection of these lines was determined as spin-glass temperature $T_{SG}$ = 60 K (see inset of Fig. 10). This transition extends down to a value of $T_f$ = 5.1 ± 0.1 K, below which the spins in the shell layer become completely frozen in a specific configuration across the surface of the sample. The linear behavior emphasized in the inset of Fig. 10, namely $H_R$ vs $T$, has been reported in the literature while associating the slope of straight line with the nanoparticle size [73,74].

Extra evidence of the spin-glass-like characteristic of the (Fe$_{0.25}$Mn$_{0.75}$)$_2$O$_3$ nanoparticles' shell layer is the upshift of the *g*-value while lowering the temperature of the sample, particularly below about 60 K (see inset of Fig. 9). It is worth mentioning that high *g*-values are characteristic of magnetically isolated transition metal ions in a low-symmetry environment, attributed to the presence of relatively strong crystalline fields due to the lack of symmetry translation, as reported in the literature [75].

## 4. CONCLUSIONS

The present study reports on the structural and magnetic characterizations of a nanostructured powder compound, which was initially labeled as $FeMnO_3$, successfully synthesized through the sol-gel method. Mössbauer spectroscopy data analysis provided evidence supporting the predominant formation of a $(Fe,Mn)_2O_3$ phase and a minority hematite ($\alpha$-$Fe_2O_3$) phase. The data extracted from the Mössbauer spectrum at 80 K were used as constraints in the Rietveld analysis of the x-ray diffraction data. The Rietveld refinement method confirmed the formation of the majority bixbyite phase (86 mol%, 94 wt%), with $(Fe_{0.25}Mn_{0.75})_2O_3$ stoichiometry and mean crystallite size of ~ 48 nm, plus the minority hematite phase (14 mol%, i.e. 6 wt%, and mean crystallite size of ~8 nm).

TEM images depict agglomerates of nanoparticles exhibiting a wide range of sizes and shapes. HRTEM images distinctly reveal crystalline planes identified as belonging to the $(Fe_{0.25}Mn_{0.75})_2O_3$ and $\alpha$-$Fe_2O_3$ phases. This observation suggests that the smaller $\alpha$-$Fe_2O_3$ nanoparticles may be decorating the larger $(Fe_{0.25}Mn_{0.75})_2O_3$ nanoparticles.

The Raman spectrum of the compound displays five Raman active modes (three Fg modes at 659 cm$^{-1}$, 519 cm$^{-1}$, and 416 cm$^{-1}$; Eg + Fg mode located at 335 cm$^{-1}$; and Eg mode at 274 cm$^{-1}$). These modes are characteristic of an $(Fe_x,Mn_{1-x})_2O_3$ phase. The spectrum did not show features that could be associated with the minority phase. On the other hand, X-ray photoelectron spectroscopy analysis confirmed the presence of oxygen vacancy onto the $(Fe_{0.25}Mn_{0.75})_2O_3$ particle surface, with varying oxidation states ($Fe^{3+}$, $Fe^{2+}$, $Mn^{3+}$, and $Mn^{4+}$).

The small size of the hematite crystallites explains why the Morin transition is not observed. At 300 K, the magnetization data indicated a dominant paramagnetic behavior credited to the $(Fe_{0.25}Mn_{0.75})_2O_3$ phase plus a weak superparamagnetic contribution coming mainly from the $\alpha$-$Fe_2O_3$ phase. Moreover, the hysteresis cycle recorded at 5 K is characteristic of ferrimagnetic ordering, revealing a phase transition associated to the $(Fe_{0.25}Mn_{0.75})_2O_3$ phase while decreasing the temperature. The ZFC peak at 33 K was attributed to the paramagnetic-ferrimagnetic transition ($T_C$), in addition to a valley observed at $T_B$ = 140 K, herein interpreted as a signature of the blocking temperature associated to the $\alpha$-$Fe_2O_3$ phase. Above Griffiths temperature ($T_G$), the Curie-Weiss law rules the temperature dependence of the susceptibility, indicating a paramagnetic phase with strong antiferromagnetic short-range correlation, with $\theta$ = -177 K and $\mu_{eff}$ = 5.2 $\mu_B$ per magnetic ion pair in the $(Fe_{0.25}Mn_{0.75})_2O_3$ phase. Importantly, the downward deviation of the inverse magnetic susceptibility ($\chi^{-1}$) versus temperature data from the ideal Curie-Weiss law, observed below $T_G$, strongly suggest the onset of the Griffiths phase (GP) in the system. Consequently, a power law using a generalized Curie-Weiss expression was used, and the value of $\lambda = 0.51$ was obtained, which is within the limit ($0 \leq \lambda < 1$) expected from a system exhibiting a GP regime. Hysteresis curve at 5 K shows a low coercive field of 4 kOe, with the magnetization not reaching saturation at 70 kOe, suggesting the occurrence of a ferrimagnetic core with a magnetic disorder at surface, characteristic of core-shell spin-glass-like behavior.

X-band magnetic resonance (MR) data revealed a strong and broad resonance line in the whole temperature range (4.3 K ≤ T ≤ 300 K), dominated by the majority phase, with $g$-value decreasing monotonically from (2.93 ± 0.01) at 50 K down to (2.18 ± 0.01) at 300 K. The temperature dependence of both resonance field and resonance linewidth shows a remarkable change in the range of 40-50 K, herein credited to surface spin glass behavior. The model picture used to explain the MR data in the lower temperature range (below about 50 K) assumes $(Fe_{0.25}Mn_{0.75})_2O_3$ nanoparticles with a core-shell structure. Results indicate that below about 50 K the shell's spin system reveals a paramagnetic to spin-glass-like transition upon cooling, with a critical temperature estimated at (43 ± 1) K. In the higher temperature range (above about 50 K), the superparamagnetic minority phase contributes remarkably to the temperature dependence of the resonance linewidth.

## Author Contributions

**John C. Mantilla** conduct all experiments and characterizations performing the data analysis. **Luiz C. C. Nagamine** performed and assisted interpretation of MS. **Daniel R. Cornejo** and **Renato Cohen** performed the magnetic measurement. **Wesley de Oliveira** helped sintering the nanocrystals and the experimentation. **Paulo Souza** performed MR measurement and helped analyzed the data. **Sebastião W. da Silva** conducted Raman spectroscopy experiment and analyzed the results. **Fermin F. H. Aragon and Pedro L. Gastelois** performed XPS measurements and analyzed the results. **Luiz. C. C. M. Nagamine**, **Paulo C. Morais, Daniel R. Cornejo and Jose A. H. Coaquira** wrote the manuscript and contributed to the structural and magnetic analysis. All authors contributed to the analysis and discussion of the results and have approved the final version of the manuscript.

## Conflicts of interest

The authors declare they have no competing financial interest or personal relationships that could have appeared to influence the work reported in this study.

## Acknowledgements

Authors acknowledge the *Laboratório Multiusuário de Microscopia de Alta Resolução* (Labmic/UFG), IF/UnB and Laboratory of Crystallography, IF/USP for XRD measurements. John C. Mantilla acknowledges partial financial support from Brazilian National Research Council (CNPq Grant # 101441/2022-3).